\documentclass{article}
\usepackage{amssymb}
\usepackage{amsfonts}
\usepackage{amsmath}

\input{tcilatex}

\begin{document}

\title{ Energy-levels crossing and radial Dirac equation: Supersymmetry and
quasi-parity spectral signatures}
\author{Omar Mustafa \\
Department of Physics, Eastern Mediterranean University, \\
G Magusa, North Cyprus, Mersin 10,Turkey\\
E-mail: omar.mustafa@emu.edu.tr}
\maketitle

\begin{abstract}
The (3+1)-dimensional Dirac equation with position dependent mass in
4-vector electromagnetic fields is considered. Using two over-simplified
examples (the Dirac-Coulomb and Dirac-oscillator fields), we report
energy-levels crossing as a spectral property or as an effect of the hidden
supersymmetric quantum mechanical language and/or quasi-parity signatures.
Under different settings of the related interactions' way-of-coupling into
Dirac equation, it is observed that the two ultimate/effective descendents,
Dirac-Coulomb and Dirac-oscillator, exhibit different conditions on the
energy-levels crossings.

\end{abstract}

\section{Introduction}

The search for exact-solvability for quantum mechanical systems, both
non-relativistic (Schr\"{o}dinger equation) and relativistic (e.g.,
Klein-Gordon and Dirac equations), is inviting and desirable.
Exactly-solvable quantum mechanical systems are vital ingredients for (among
others that are mathematically motivated, say) the conceptual understanding
of physics and for inspiring the structure of the numerical methods designed
to solve more complicated physical problems. However, whilst intensive
attention was paid to exact-solvability of the non-relativistic Schr\"{o}%
dinger equation, the relativistic Klein-Gordon and Dirac equations remained
unfortunate and only partially attended in the yet already
partially-explored Dirac territories.

Some exactly-solvable potentials, for example, are known to belong to some
distinctive classes of shape invariant potentials [1]. Within each class of
which the so-called point-canonical-transformation (PCT) [2] would map the
solution (eigenvalues and eigenfunctions) of one into another. On the other
hand, supersymmetric quantum mechanics [3] and potential algebras [4] (among
others of course) are known to be used to obtain exact solutions for quantum
mechanical systems. Yet, in between exactly-solvable and
non-exactly-solvable there exists the gray zone of
conditionally-exactly-solvable (i.e., all the spectrum is obtained) [5] and
quasi-exactly-solvable (i.e., part of the spectrum is obtained) [6]
potentials models.

Nevertheless, we may recollect that the supersymmetric quantum mechanical
language is realized as a \emph{hidden/built-in symmetry} in the
(1+1)-dimensional Dirac equation (cf., e.g. [3,7--9]). Nogami and Toyama [7]
have reported that the associated Schr\"{o}dinger supersymmetric-partner
Hamiltonians share the same energy spectrum including the lowest states
unless Dirac equation allows a zero-mode (i.e., zero-energy bound-state).
Jackiw and Rebbi [9] have, moreover, reported that only for some certain 
\emph{topological }trends where the Lorentz scalar potential $S(x)$ is
localized (i.e., $S(x)\rightarrow 0$ for $x\rightarrow \pm \infty $), Dirac
equation would allow a zero-mode.

One should be reminded, hereupon, that a Lorentz scalar potential (and/or an
almost mathematically and partially-physically equivalent position-dependent
mass) in Dirac equation is mainly motivated by the MIT bag model of quarks
(cf., e.g., [10] and references therein). Yet, a Lorentz scalar potential
couples to the mass of the fermion instead of its charge and the related
positive and negative energies exhibit identical behaviors. Moreover, the
position-dependent mass settings are useful models to study, for example,
the energy density many-body problem, electronic properties of
semiconductors and quantum dots, etc (cf., e.g., sample of references in
[11-18]). We may also recollect that the phenomenon of \emph{energy-levels
crossing }is responsible for electron transfer in protein, it underlies
stability analysis in mechanical engineering, and mathematically appears in
algebraic geometry (cf., e.g., [19] and related references therein).

Very recently [20], we have reemphasized the hidden/built-in supersymmetric
quantum mechanical language in the spectrum of the (1+1)-dimensional Dirac
equation with position-dependent mass and complexified Lorentz scalar
interactions. We have reported the "quasi-parity" signature on the Dirac
spectrum and discussed energy-levels crossings related to supersymmetry
and/or "quasi-parity" signatures. We have observed that the supersymmetric
signature on the (1+1)-Dirac spectrum is documented through the emergence of
"exact" isospectral (i.e., including the lowest-state) partner Hamiltonians
for "even"-quasi-parity, whereas the partner Hamiltonians share the same
energy spectrum with a "missing" lowest-state for "odd"-quasi-parity, at
least for the examples discussed therein. It would be interesting, we
contemplate, if such studies are extended to cover the (3+1)-dimensional
radial Dirac equation with different models of interactions' couplings and
with position-dependent mass settings. Such studies merely exist in the
literature, to the best of our knowledge, and may very well add a new
flavour to the readily "multi-flavoured" Dirac equation.

This article is organized as follows. In section 2, we recollect the
(3+1)-dimensional radial Dirac equation with position-dependent mass in a
four-vector electromagnetic field. Therein, we realize that the decoupled
one-dimensional Schr\"{o}dinger-like radial Dirac equations exhibit
supersymmetric language only when $M\left( r\right) =V\left( r\right) =0$,
where $M\left( r\right) =m\left( r\right) +S\left( r\right) $ with the
position-dependent mass $m\left( r\right) $, the Lorentz scalar potential $%
S\left( r\right) $, and the Lorentz vector potential $V\left( r\right) $. In
section 3, the consequences of an equally-mixed non-zero Lorentz vector and
Lorentz scalar potential settings are discussed through illustrative
examples: Dirac-Coulomb-I, Dirac-oscillator-I, and Dirac-oscillator-II. The
consequences of $V\left( r\right) \neq M\left( r\right) $ with the magnetic
interaction $A\left( r\right) =-$ $\zeta _{2}^{\prime }\left( r\right) /%
\left[ 2\zeta _{2}\left( r\right) \right] \neq 0$ (see Eq.(7) below) are
given in section 4, along with illustrative examples:
Dirac/Klein-Gordon-Coulomb-II and Dirac/Klein-Gordon-oscillator-III. In
section 5, we report the consequences of an equally-mixed Lorentz vector and
scalar \emph{"free"}-fields (i.e., $M\left( r\right) =V\left( r\right) =0$).
The supersymmetric quantum mechanical language and the "quasi-parity"
signatures on the spectra of a \emph{Dirac-oscillator-toy} and a \emph{%
Dirac-Coulomb-toy} models are reported in the same section. We conclude in
section 6.

\section{Radial Dirac equation with position dependent mass in a 4-vector
electromagnetic field, recollected}

The Hamiltonian describing a Dirac particle (in $\hbar =c=e=1$ units) in a
four-vector electromagnetic field $A_{\mu }=\left( A_{0},\vec{A}\right)
=\left( V,\vec{A}\right) $ reads%
\begin{equation}
H_{D}=\overrightarrow{\alpha }\cdot \left( \overrightarrow{p}-i%
\overrightarrow{A}\right) +\beta m+V,
\end{equation}%
with%
\begin{equation*}
\alpha _{j}=\left( 
\begin{array}{cc}
0 & \sigma _{j} \\ 
\sigma _{j} & 0%
\end{array}%
\right) \text{ },\text{ \ }\beta =\left( 
\begin{array}{cc}
1 & 0 \\ 
0 & -1%
\end{array}%
\right) ,
\end{equation*}%
where $\sigma _{j}$ are Pauli's $2\times 2$ matrices and $1$ \ is the $%
2\times 2$ unit matrix. Under spherically symmetric settings, $\vec{A}%
\rightarrow \hat{r}A\left( r\right) $, $V\rightarrow V\left( r\right) $
accompanied (for the convenience of the current study) by a
position-dependent mass Lorentz scalar field, $m\rightarrow m+m\left(
r\right) +S\left( r\right) =m+M\left( r\right) $, the two-component Dirac
equation reads%
\begin{equation}
\left( 
\begin{array}{cc}
m+M\left( r\right) +V\left( r\right) & \frac{\kappa }{r}+A\left( r\right)
-\partial _{r}\medskip \\ 
\frac{\kappa }{r}+A\left( r\right) +\partial _{r} & -m-M\left( r\right)
+V\left( r\right)%
\end{array}%
\right) \left( 
\begin{array}{c}
g\left( r\right) \medskip \\ 
f\left( r\right)%
\end{array}%
\right) =E\left( 
\begin{array}{c}
g\left( r\right) \medskip \\ 
f\left( r\right)%
\end{array}%
\right)
\end{equation}%
with energy $E$, and $\kappa $ in the centrifugal term is given by%
\begin{equation}
\kappa =\left\{ 
\begin{tabular}{ll}
$-\left( \ell +1\right) $ & for $j=\ell +1/2\medskip $ \\ 
$\ell $ & for \ $j=\ell -1/2$%
\end{tabular}%
\right. \,\,\Longrightarrow \,\,\kappa \left( \kappa +1\right) =\ell \left(
\ell +1\right) ,
\end{equation}%
where $\ell =0,1,2,\cdots $ is the angular momentum quantum number. Equation
(2) decouples into%
\begin{equation}
\zeta _{1}\left( r\right) \,g\left( r\right) -\left( \frac{\kappa }{r}%
+A\left( r\right) -\partial _{r}\right) \,f\left( r\right) =0
\end{equation}%
\begin{equation}
\zeta _{2}\left( r\right) \,f\left( r\right) -\left( \frac{\kappa }{r}%
+A\left( r\right) +\partial _{r}\right) \,g\left( r\right) =0
\end{equation}%
where%
\begin{equation}
\zeta _{1}\left( r\right) =\left( E-m\right) -V\left( r\right) -M\left(
r\right) ,
\end{equation}%
\begin{equation}
\zeta _{2}\left( r\right) =\left( E+m\right) -V\left( r\right) +M\left(
r\right) .
\end{equation}%
Substituting $f\left( r\right) $ of (5) into (4) would, with%
\begin{equation}
\tilde{A}\left( r\right) =\frac{\kappa }{r}+A\left( r\right) ,
\end{equation}%
imply%
\begin{equation}
\left\{ -\partial _{r}^{2}+\tilde{A}\left( r\right) ^{2}-\tilde{A}^{\prime
}\left( r\right) +\frac{\zeta _{2}^{\prime }\left( r\right) }{\zeta
_{2}\left( r\right) }\left[ \tilde{A}\left( r\right) +\partial _{r}\right]
-\zeta _{1}\left( r\right) \zeta _{2}\left( r\right) \right\} g\left(
r\right) =0
\end{equation}%
where primes denote derivatives with respect to $r$. Moreover, a
substitution of the form%
\begin{equation}
g\left( r\right) =\phi _{2}\left( r\right) \exp \left( -\frac{P_{2}\left(
r\right) }{2}\right) \text{ ; \ }P_{2}^{\prime }\left( r\right) =\frac{%
V^{\prime }\left( r\right) -M^{\prime }\left( r\right) }{\zeta _{2}\left(
r\right) }\,
\end{equation}%
would remove the first order derivative and result in a one-dimensional Schr%
\"{o}dinger-like equation%
\begin{equation}
\left\{ -\partial _{r}^{2}+\tilde{A}\left( r\right) ^{2}-\tilde{A}^{\prime
}\left( r\right) +U_{2}\left( r\right) -\zeta _{1}\left( r\right) \zeta
_{2}\left( r\right) \right\} \phi _{2}\left( r\right) =0
\end{equation}%
where%
\begin{equation}
U_{2}\left( r\right) =\frac{\zeta _{2}^{\prime }\left( r\right) }{\zeta
_{2}\left( r\right) }\tilde{A}\left( r\right) +\left[ \frac{3}{4}\left( 
\frac{\zeta _{2}^{\prime }\left( r\right) }{\zeta _{2}\left( r\right) }%
\right) ^{2}-\frac{1}{2}\frac{\zeta _{2}^{\prime \prime }\left( r\right) }{%
\zeta _{2}\left( r\right) }\right] .
\end{equation}%
Similarly, substituting $g\left( r\right) $ of (4) into (5) and taking%
\begin{equation*}
f\left( r\right) =\,\phi _{1}\left( r\right) \exp \left( -\frac{P_{1}\left(
r\right) }{2}\right) \text{ ; \ }P_{1}^{\prime }\left( r\right) =\frac{%
V^{\prime }\left( r\right) +M^{\prime }\left( r\right) }{\zeta _{1}\left(
r\right) }
\end{equation*}%
would imply%
\begin{equation}
\left\{ -\partial _{r}^{2}+\tilde{A}\left( r\right) ^{2}+\tilde{A}^{\prime
}\left( r\right) +U_{1}\left( r\right) -\zeta _{1}\left( r\right) \zeta
_{2}\left( r\right) \right\} \phi _{1}\left( r\right) =0
\end{equation}%
where%
\begin{equation}
U_{1}\left( r\right) =-\frac{\zeta _{1}^{\prime }\left( r\right) }{\zeta
_{1}\left( r\right) }\tilde{A}\left( r\right) +\left[ \frac{3}{4}\left( 
\frac{\zeta _{1}^{\prime }\left( r\right) }{\zeta _{1}\left( r\right) }%
\right) ^{2}-\frac{1}{2}\frac{\zeta _{1}^{\prime \prime }\left( r\right) }{%
\zeta _{1}\left( r\right) }\right] .
\end{equation}%
It should be noted that the decoupled radial Dirac one-dimensional Schr\"{o}%
dinger-like equations, (11) and (13), possess natural (though
hidden/built-in) supersymmetric quantum mechanical language only when $%
M\left( r\right) =V\left( r\right) =0$. In the forthcoming experiments we
shall be focusing on the upper component one-dimensional Schr\"{o}%
dinger-like radial Dirac equation in (11) and (12).

\section{Consequences of an equally-mixed Lorentz vector and scalar fields; $%
V\left( r\right) =M\left( r\right) $}

An equally-mixed Lorentz vector and scalar fields, $V\left( r\right)
=M\left( r\right) $, would imply $\zeta _{2}\left( r\right) =E+m$, $\zeta
_{1}\left( r\right) =E-m-2M\left( r\right) $. Consequently, (11) reduces to%
\begin{equation}
\left\{ -\partial _{r}^{2}+\tilde{A}\left( r\right) ^{2}-\tilde{A}^{\prime
}\left( r\right) +2M\left( r\right) \left[ E+m\right] \right\} \phi
_{2}\left( r\right) =\left[ E^{2}-m^{2}\right] \phi _{2}\left( r\right) ,
\end{equation}%
where $\tilde{A}\left( r\right) =\frac{\kappa }{r}+A\left( r\right) $, as
given in (8)$.$

\subsection{Dirac-Coulomb-I:}

For $A\left( r\right) =a/r$ and $M\left( r\right) =b/r$ equation (15) reads%
\begin{equation}
\left\{ -\partial _{r}^{2}+\frac{\tilde{\kappa}\left( \tilde{\kappa}%
+1\right) }{r^{2}}+\frac{2b\left[ E+m\right] }{r}\right\} \phi _{2}\left(
r\right) =\left[ E^{2}-m^{2}\right] \phi _{2}\left( r\right) ,
\end{equation}%
where%
\begin{equation*}
\tilde{\kappa}=\kappa +a=\left\{ 
\begin{tabular}{ccl}
$a\medskip +\left( j+1/2\right) $ & for & $\kappa =+\left( j+1/2\right) $ \\ 
$a\medskip -\left( j+1/2\right) $ & for & $\kappa =-\left( j+1/2\right) $%
\end{tabular}%
\right. .
\end{equation*}%
Equation (16) admits exact solution that can be very well inferred from the
well known radial Schr\"{o}dinger-Coulomb problem to yield (with the radial
quantum number $n_{r}=0,1,2,\cdots $)%
\begin{equation}
\left[ E^{2}-m^{2}\right] \medskip =-\frac{b^{2}\left[ E+m\right] ^{2}}{%
\tilde{n}^{2}};\,\,\tilde{n}=n_{r}+\tilde{\kappa}+1>0,
\end{equation}%
which in turn implies%
\begin{equation}
E=\frac{m\left( \tilde{n}^{2}-b^{2}\right) }{\tilde{n}^{2}+b^{2}}.\medskip
\allowbreak
\end{equation}%
However, the fact that $\tilde{\kappa}=a\pm \left( j+1/2\right) $ would
manifest energy-levels crossings to obtain. That is, a state labeled by $\,%
\tilde{n}_{1}=n_{r1}+a+j_{1}+3/2$ would cross with a state labeled by $%
\tilde{n}_{2}=n_{r2}+a-j_{2}+1/2$ when%
\begin{equation*}
\frac{\tilde{n}_{1}^{2}-b^{2}}{\tilde{n}_{1}^{2}+b^{2}}=\frac{\tilde{n}%
_{2}^{2}-b^{2}}{\tilde{n}_{2}^{2}+b^{2}}\Longrightarrow \tilde{n}_{2}=\tilde{%
n}_{1}\Longrightarrow n_{r2}-n_{r1}=j_{1}+j_{2}+1.
\end{equation*}%
Moreover, one may wish to mind the consequences associated with a
complexified coupling constant in $M\left( r\right) $, and hence in $V\left(
r\right) $, in such a way that $V\left( r\right) =M\left( r\right)
=b/r=-ib_{\circ }/r$ (e.g., simulating, say, the interaction of a point
nucleus with an imaginary charge $iZe$ and a particle of charge $-e$). In
this case, $V\left( r\right) =M\left( r\right) $ is $\mathcal{PT}$%
-symmetrized [21] and%
\begin{equation*}
E_{\mathcal{PT}}=\frac{m\left( \tilde{n}^{2}+b_{\circ }^{2}\right) }{\tilde{n%
}^{2}-b_{\circ }^{2}}\allowbreak
\end{equation*}%
Not only the energy spectrum $E_{\mathcal{PT}}$ \ follows similar
energy-levels crossing scenario as that of (18), but also it suffers from
the so called \emph{flown-away} (cf., e.g., [21]) states that disappear \
from the spectrum when $\tilde{n}=\left\vert b_{\circ }\right\vert .$

\subsection{Dirac-Oscillator-I:}

For $A\left( r\right) =br/2$ and $M\left( r\right) =0$ equation (15) implies 
\begin{equation}
\left\{ -\partial _{r}^{2}+\frac{\ell \left( \ell +1\right) }{r^{2}}+\frac{%
b^{2}}{4}r^{2}+\kappa b-\frac{b}{2}\right\} \phi _{2}\left( r\right) =\left[
E^{2}-m^{2}\right] \phi _{2}\left( r\right) ,
\end{equation}%
where $\,\kappa \left( \kappa +1\right) =\ell \left( \ell +1\right) $ for
both $\kappa =-\left( \ell +1\right) ;j=\ell +1/2$ and $\kappa =\ell ;j=\ell
-1/2$ is considered. This would result in%
\begin{equation}
E^{2}-m^{2}=b\left( 2n_{r}+\ell +3/2\right) +\kappa b-\frac{b}{2},
\end{equation}%
to imply%
\begin{equation}
E_{\pm }=\medskip \pm \sqrt{m^{2}+b\left( 2n_{r}+\ell +\kappa +1\right) }.
\end{equation}%
Obviously this result depends on the combination of the quantum numbers $%
2n_{r}+\ell =\Lambda $ and splits into%
\begin{equation}
E_{\pm }=\left\{ 
\begin{tabular}{ccc}
$\pm \sqrt{m^{2}+b\left( \Lambda +j+3/2\right) }$ & \medskip for & $\kappa
=+\left( j+1/2\right) $ \\ 
$\pm \sqrt{m^{2}+b\left( \Lambda -j-1/2\right) }$ & \medskip for & $\kappa
=-\left( j+1/2\right) $%
\end{tabular}%
\right.
\end{equation}%
One should pay attention to the possible energy-levels crossings that occur
between positive energy sets or negative energy sets. These are unavoidable
energy-levels crossings manifested by $\kappa =\pm \left( j+1/2\right) $ and
admits the following scenario: A state labeled by $\Lambda _{1}$ and $j_{1}$
crosses with a state labeled by $\Lambda _{2}$ and $j_{2}$ for all $b$
values when%
\begin{equation*}
\Lambda _{1}+j_{1}+\frac{3}{2}=\Lambda _{2}-j_{2}-\frac{1}{2}\Longrightarrow
\Lambda _{2}-\Lambda _{1}=j_{1}+j_{2}+2
\end{equation*}%
Nevertheless, quantum numbers related degeneracies are also feasible at
different values of $b$. That is, when%
\begin{equation*}
\Lambda _{1}=\Lambda _{2}=\Lambda \text{, }j_{1}\neq j_{2}\Longrightarrow
\Lambda \left( b_{2}-b_{1}\right) =b_{1}j_{1}+b_{2}j_{2}+\frac{3b_{1}+b_{2}}{%
2}
\end{equation*}%
and when%
\begin{equation*}
\Lambda _{1}\neq \Lambda _{2}\text{, }j_{1}=j_{2}=j\Longrightarrow \Lambda
_{2}b_{2}-\Lambda _{1}b_{1}=j\left( b_{1}+b_{2}\right) +\frac{3b_{1}+b_{2}}{2%
}
\end{equation*}

\subsection{Dirac-Oscillator-II:}

For $A\left( r\right) =a/r$ and $M\left( r\right) =B^{2}r^{2}/2$ equation
(15) reads%
\begin{equation}
\left\{ -\partial _{r}^{2}+\frac{\tilde{\kappa}\left( \tilde{\kappa}%
+1\right) }{r^{2}}+B^{2}\left[ E+m\right] \,r^{2}\right\} \phi _{2}\left(
r\right) =\left[ E^{2}-m^{2}\right] \phi _{2}\left( r\right) ,
\end{equation}%
In this case%
\begin{equation}
\left[ E^{2}-m^{2}\right] =2B\sqrt{E+m}\left( 2n_{r}+\tilde{\kappa}%
+3/2\right) \medskip
\end{equation}%
which would lead to $E=-m$ (to be discarded) and, with $\tilde{N}\medskip
_{\pm }=2n_{r}+a\pm \left( j+1/2\right) +3/2>0,$%
\begin{equation}
E=-m+\left( \frac{1}{3}\xi _{\pm }^{1/3}+2m\xi _{\pm }^{-1/3}\right) ^{2}
\end{equation}%
where%
\begin{equation}
\text{\ }\xi _{\pm }=27B\,\tilde{N}_{\pm }+3\sqrt{-24m^{3}+81B^{2}\tilde{N}%
_{\pm }^{2}};\text{ \ }\tilde{N}_{\pm }\geq \sqrt{\frac{8m^{3}}{27B^{2}}}%
\text{\ }
\end{equation}%
In this case, it is obvious that energy-levels crossings occur when \ $\xi
_{+}=$\ $\xi _{-}$. One may, for the sake of simplicity, choose the case
where%
\begin{equation*}
\tilde{N}_{\pm }=\sqrt{\frac{8m^{3}}{27B^{2}}}\Longrightarrow \text{\ }\xi
_{\pm }=27B\,\tilde{N}_{\pm },
\end{equation*}%
which would imply that a state labeled by $n_{r1}$ and $j_{1}$ crosses with
a state labeled by $n_{r2}$ and $j_{2}$, for all $a$, when%
\begin{equation*}
\xi _{+}=\ \xi _{-}\Longrightarrow \tilde{N}_{+}=\ \tilde{N}%
_{-}\Longrightarrow n_{r2}-n_{r1}=\left( j_{1}+j_{2}+1\right) /2
\end{equation*}

\section{Consequences of $A\left( r\right) =-$ $\protect\zeta _{2}^{\prime
}\left( r\right) /\left[ 2\protect\zeta _{2}\left( r\right) \right] \neq 0$
and $V\left( r\right) \neq M\left( r\right) $}

If we consider the class of interactions where $A\left( r\right) =-$ $\zeta
_{2}^{\prime }\left( r\right) /2\zeta _{2}\left( r\right) \neq 0$ and $%
V\left( r\right) \neq M\left( r\right) $, Dirac equation in (11) reduces to%
\begin{equation}
\left\{ -\partial _{r}^{2}+\frac{\kappa \left( \kappa +1\right) }{r^{2}}%
-\zeta _{1}\left( r\right) \zeta _{2}\left( r\right) \right\} \phi
_{2}\left( r\right) =0,
\end{equation}%
which is, in fact, in exact form as that of the radial one-dimensional
Klein-Gordon (KG) equation%
\begin{equation}
\left\{ -\partial _{r}^{2}+\frac{\ell \left( \ell +1\right) }{r^{2}}-\left[
E-V\left( r\right) \right] ^{2}+\left[ m+M\left( r\right) \right]
^{2}\right\} \phi _{2}\left( r\right) =0.
\end{equation}%
Moreover, it should be noted hereby that the case where $A\left( r\right)
=-\kappa /r-$ $\zeta _{2}^{\prime }\left( r\right) /2\zeta _{2}\left(
r\right) \neq 0$ and $V\left( r\right) \neq M\left( r\right) $ corresponds
to the $s$-waves (i.e., $\ell =0$) solution of (28). Hence, one need not
consider it as a separate case to deal with.

\subsection{Dirac/Klein-Gordon-Coulomb-II:}

For $V(r)=\alpha _{1}/r$ and $M\left( r\right) =\alpha _{2}/r$ equation (28)
reads%
\begin{equation}
\left\{ -\partial _{r}^{2}+\frac{\mathcal{L}\left( \mathcal{L}+1\right) }{%
r^{2}}+\frac{2\left[ \alpha _{1}E+\alpha _{2}m\right] }{r}\right\} \phi
_{2}\left( r\right) =\left[ E^{2}-m^{2}\right] \phi _{2}\left( r\right) ,
\end{equation}%
with%
\begin{equation}
\mathcal{L=-}\frac{1}{2}+\sqrt{\left( \ell +1/2\right) ^{2}-\alpha
_{1}^{2}+\alpha _{2}^{2}}\geq 0,
\end{equation}%
and admits exact solution of the form%
\begin{equation*}
\left[ E^{2}-m^{2}\right] =-\frac{\left[ \alpha _{1}E+\alpha _{2}m\right]
^{2}}{\mathcal{N}^{2}};\,\mathcal{N}=n_{r}+\mathcal{L}+1>0.
\end{equation*}%
This would lead to%
\begin{equation}
\frac{E_{\pm }}{m}=\frac{-\alpha _{1}\alpha _{2}}{\mathcal{N}^{2}+\alpha
_{1}^{2}}\pm \left[ \left( \frac{\alpha _{1}\alpha _{2}}{\mathcal{N}%
^{2}+\alpha _{1}^{2}}\right) ^{2}+\frac{\mathcal{N}^{2}-\alpha _{2}^{2}}{%
\mathcal{N}^{2}+\alpha _{1}^{2}}\right] ^{1/2},
\end{equation}%
to yield, for various especial cases of coupling constants,%
\begin{equation}
\frac{E_{\pm }}{m}=\left\{ 
\begin{tabular}{ccc}
$\pm \left[ 1-\alpha _{2}^{2}/\mathcal{N}^{2}\right] ^{1/2}\medskip ;$ $%
\mathcal{N\geq }\left\vert \alpha _{2}\right\vert $ & for & $\alpha
_{1}=0\wedge \alpha _{2}\neq 0$ \\ 
$\pm \left[ 1+\alpha _{1}^{2}/\mathcal{N}^{2}\right] ^{-1/2}\medskip $ & for
& $\alpha _{1}\neq 0\wedge \alpha _{2}=0$ \\ 
$\left( -\alpha _{\circ }^{2}\pm \mathcal{N}^{2}\right) /\left( \mathcal{N}%
^{2}+\alpha _{\circ }^{2}\right) \medskip $ & for & $\alpha _{1}=\alpha
_{2}=\alpha _{\circ }\neq 0$%
\end{tabular}%
\right.
\end{equation}%
It should be obvious that energy-levels crossings for (32) are not feasible
at all. At this point, one should notice that the effect of
positive/negative $\kappa $ is absent in the process of choosing $A\left(
r\right) =-$ $\zeta _{2}^{\prime }\left( r\right) /2\zeta _{2}\left(
r\right) \neq 0$ and $V\left( r\right) \neq M\left( r\right) $. However, the
complexification of the coupling constants would manifest \emph{flown-away}
states (cf., e.g., [21] ) to obtain and disappear from the spectrum. For
example, for the case%
\begin{equation*}
\alpha _{1}=\alpha _{2}=i\alpha _{\circ }\neq 0\Longrightarrow \frac{E_{\pm }%
}{m}=\frac{\alpha _{\circ }^{2}\pm \mathcal{N}^{2}}{\mathcal{N}^{2}-\alpha
_{\circ }^{2}}\medskip ,
\end{equation*}%
the energy states \emph{fly-away} and disappear from the spectrum when $%
\mathcal{N=}\left\vert \alpha _{\circ }\right\vert $, and for%
\begin{equation*}
\alpha _{1}=i\alpha _{1}^{\prime }\neq 0,\alpha _{2}=0\Longrightarrow \frac{%
E_{\pm }}{m}=\pm \sqrt{\frac{\mathcal{N}^{2}}{\mathcal{N}^{2}-\alpha
_{1}^{\prime 2}}}\medskip ,
\end{equation*}%
energy states \emph{fly-away} when $\mathcal{N=}\left\vert \alpha
_{1}^{\prime }\right\vert $. However, \emph{flown-away} states never occur
for the case%
\begin{equation*}
\alpha _{1}=0,\alpha _{2}=i\alpha _{2}^{\prime }\Longrightarrow \frac{E_{\pm
}}{m}=\pm \sqrt{\frac{\mathcal{N}^{2}-\alpha _{2}^{\prime 2}}{\mathcal{N}^{2}%
}},
\end{equation*}%
but rather the system loses its observability (i.e., reality and
discreteness) and collapses when $\left\vert \alpha _{2}^{\prime
}\right\vert >\mathcal{N}$. Nevertheless, in connection with such
complexified version of $\alpha _{1}$ and $\alpha _{2}$, the reader may
consult Mustafa [21] for more comprehensive details on the spectral
properties of a similar complexified Coulombic fields (but with $\gamma $ in
[21] replacing $\mathcal{L}$ in the current example).

\subsection{Dirac/Klein-Gordon-Oscillator-III:}

For $V\left( r\right) =0$, and $M\left( r\right) =\beta _{1}/r+$ $\beta
_{2}r-m$ equation (28) implies%
\begin{equation*}
\left\{ -\partial _{r}^{2}+\frac{\mathcal{\tilde{L}}\left( \mathcal{\tilde{L}%
}+1\right) }{r^{2}}+\beta _{2}^{2}r^{2}+2\beta _{1}\beta _{2}\right\} \phi
_{2}\left( r\right) =E^{2}\phi _{2}\left( r\right) ,
\end{equation*}%
where%
\begin{equation}
E_{\pm }=\pm \sqrt{\beta _{2}\left( 4n_{r}+2\mathcal{\tilde{L}}+3\right)
+2\beta _{1}\beta _{2}}
\end{equation}%
with%
\begin{equation}
\mathcal{\tilde{L}=-}\frac{1}{2}+\sqrt{\left( \ell +1/2\right) ^{2}+\beta
_{1}^{2}}\geq 0
\end{equation}%
Obviously, neither energy-levels crossings nor \emph{flown-away} states are
feasible for this model. Nevertheless, one should pay attention to the
parametric settings that may lead to imaginary energies and consequently
system collapse.

\section{Consequences of equally-mixed "\emph{free}"-fields: supersymmetry
and quasi-parity}

It is obvious that with equally-mixed Lorentz vector and Lorentz scalar
"free"-fields (i.e., $V\left( r\right) =M\left( r\right) =0$) one would
obtain, from (11) and (13),%
\begin{equation}
\left\{ -\partial _{r}^{2}+V_{\pm }\left( r\right) \right\} \phi _{\pm
}\left( r\right) =\lambda _{\pm }\phi _{\pm }\left( r\right) ,
\end{equation}%
where $\lambda _{\pm }=E^{2}-m^{2}$, $\phi _{+}\left( r\right) =\phi
_{1}\left( r\right) $, $\phi _{-}\left( r\right) =\phi _{2}\left( r\right) $%
, and%
\begin{equation*}
V_{\pm }\left( r\right) =V_{\omega }\left( r\right) =\tilde{A}\left(
r\right) ^{2}+\omega \tilde{A}^{\prime }\left( r\right) ;\text{ \ }\omega
=\pm 1
\end{equation*}%
stands for an effective supersymmetric-like partner potentials. The
supersymmetric language is very well pronounced in this process, therefore.
Yet, the \emph{quasi-parity} shall emerge in the forthcoming two Dirac-toy
models.

\subsection{A Dirac-oscillator-toy}

Let us consider a "toy" model:%
\begin{equation*}
\tilde{A}\left( r\right) =-\frac{A}{r}+\frac{1}{2}Br;\text{ }%
\mathbb{R}
\ni A,B>0,
\end{equation*}%
that results in an effective supersymmetric partner "\emph{%
Dirac-oscillator-toy}" potential of the form%
\begin{equation}
V_{\omega }\left( r\right) =\frac{A\left( A+\omega \right) }{r^{2}}+\frac{1}{%
4}B^{2}r^{2}-B\left( A-\frac{\omega }{2}\right) .
\end{equation}%
In the repulsive/attractive-like core, moreover, one may replace $A\left(
A+\omega \right) $ by $\sigma \left( \sigma +1\right) $. In this case, $%
\sigma $ would denote quasi-angular momentum quantum number and $\sigma
=-1,0 $ may very well correspond to "even" and "odd" \emph{quasi-parity}
(i.e., $q=\left( -1\right) ^{\sigma +1}$), respectively. For more details on
quasi-parity convention the reader may refer to, e.g., Mustafa and Znojil
[22] and Znojil [23] and related references cited therein. In a
straightforward manner, however, one can show that%
\begin{equation}
\sigma \left( \sigma +1\right) =A\left( A+\omega \right) \Longrightarrow
\sigma =-\frac{1}{2}+q\left( A+\frac{\omega }{2}\right) \text{.}
\end{equation}%
Under such settings, it is obvious that both supersymmetric-like partner
potentials in (36) admit exact closed form solutions (cf., e.g., Mustafa and
Znojil [22], and Znojil [23]):%
\begin{equation}
\lambda _{\omega }=\frac{B}{2}\left( 4n_{r}+2qA+\omega q+2\right) -B\left( A-%
\frac{\omega }{2}\right) ;\text{ \ }n_{r}=0,1,2,\cdots ,\medskip
\end{equation}%
which would split (with $\omega =\pm 1$) into%
\begin{eqnarray}
\lambda _{+,q} &=&\frac{B}{2}\left( 4n_{r}+2qA+q+3\right) -BA\text{ }%
\Rightarrow \left\{ 
\begin{tabular}{l}
$\lambda _{+,q=+1}=2B\left( n_{r}+1\right) \medskip $ \\ 
$\lambda _{+,q=-1}=2B\left( n_{r}-A+\frac{1}{2}\right) \medskip $%
\end{tabular}%
\right. ,\medskip \medskip \medskip \\
\lambda _{-,q} &=&\frac{B}{2}\left( 4n_{r}+2qA-q+1\right) -BA\text{ }%
\Rightarrow \left\{ 
\begin{tabular}{l}
$\lambda _{-,q=+1}=2Bn_{r}$ \\ 
$\lambda _{-,q=-1}=2B\left( n_{r}-A+\frac{1}{2}\right) \medskip $%
\end{tabular}%
\right. .\medskip \medskip
\end{eqnarray}%
We may now pay attention to the supersymmetric language "signature" which is
documented in the facts that $\lambda _{+,q=-1}=\lambda _{-,q=-1}$, for odd
quasi-parity, and $\lambda _{+,q=+1}=\lambda _{-,q=+1}+const.=\lambda
_{-,q=+1}+2B$, for even quasi-parity. That is, for an even quasi-parity the
superpartner potentials possess identical spectra with a missing lowest
state, whereas for an odd quasi-parity the superpartner potentials are
"exactly" isospectral. Nevertheless, the consequence of such \emph{"hidden"}%
-supersymmetry in Dirac equation is very well pronounced in the related
Dirac spectra (with $\lambda _{\pm }=E^{2}-m^{2}$):%
\begin{eqnarray}
E_{+,q} &=&\left\{ 
\begin{tabular}{l}
$E_{+,q=+1}=+\sqrt{m^{2}+2B\left( n_{r}+1\right) }\medskip $ \\ 
$E_{+,q=-1}=+\sqrt{m^{2}+2B\left( n_{r}-A+\frac{1}{2}\right) }\medskip $%
\end{tabular}%
\right. \text{ };\text{ }n_{r}=0,1,2,\cdots , \\
E_{-,q} &=&\left\{ 
\begin{tabular}{l}
$E_{-,q=+1}=-\sqrt{m^{2}+2Bn_{r}}\medskip $ \\ 
$E_{-,q=-1}=-\sqrt{m^{2}+2B\left( n_{r}-A+\frac{1}{2}\right) }\medskip $%
\end{tabular}%
\right. \text{ };\text{ }n_{r}=0,1,2,\cdots .
\end{eqnarray}%
In this case, the quasi-parity signature appears in the energy-level
crossings between the two sets of energies in (41) and between those in
(42). That is, the two sets of energies in (41) cross with each other when%
\begin{equation}
E_{+}\left( n_{r}=n_{r1},q=+1\right) =E_{+}\left( n_{r}=n_{r2},q=-1\right)
\Longrightarrow n_{r2}-n_{r1}=A+\frac{1}{2}
\end{equation}%
and those in (42) cross with each other when%
\begin{equation}
E_{-}\left( n_{r}=n_{r3},q=+1\right) =E_{-}\left( n_{r}=n_{r4},q=-1\right)
\Longrightarrow n_{r4}-n_{r3}=A-\frac{1}{2}.
\end{equation}

\subsection{A Dirac-Coulomb-toy}

On the other hand, a "toy" model of the form%
\begin{equation*}
\tilde{A}\left( r\right) =-\frac{A}{r}+B;\text{ }%
\mathbb{R}
\ni A,B>0,
\end{equation*}%
would result in an effective supersymmetric partner "\emph{Dirac-Coulomb-toy}%
" potentials%
\begin{equation}
V_{\omega }\left( r\right) =\frac{A\left( A+\omega \right) }{r^{2}}-\frac{2AB%
}{r}+B^{2},
\end{equation}%
which admits exact solution of the form%
\begin{equation}
\lambda _{\omega }=-\frac{\left( AB\right) ^{2}}{\breve{n}^{2}}+B^{2};\ 
\breve{n}=n_{r}+\sigma +1>0.\text{ }n_{r}=0,1,2,\cdots .\medskip
\end{equation}%
This result would split into%
\begin{eqnarray}
\lambda _{+,q} &=&B^{2}\left\{ 1-A^{2}\left[ n_{r}+q\left( A+\frac{1}{2}%
\right) +\frac{1}{2}\right] ^{-2}\right\} \text{ },\medskip \medskip \medskip
\\
\lambda _{-,q} &=&B^{2}\left\{ 1-A^{2}\left[ n_{r}+q\left( A-\frac{1}{2}%
\right) +\frac{1}{2}\right] ^{-2}\right\} .\medskip
\end{eqnarray}%
Each of which, respectively, yields%
\begin{equation}
\lambda _{+,q}\Longrightarrow \left\{ 
\begin{tabular}{l}
$\lambda _{+,q=+1}=B^{2}\left\{ 1-A^{2}\left[ n_{r}+A+1\right] ^{-2}\right\}
\medskip $ \\ 
$\lambda _{+,q=-1}=B^{2}\left\{ 1-A^{2}\left[ n_{r}-A\right] ^{-2}\right\}
\medskip $%
\end{tabular}%
\right.
\end{equation}%
\begin{equation}
\lambda _{-,q}\Longrightarrow \left\{ 
\begin{tabular}{l}
$\lambda _{-,q=+1}=B^{2}\left\{ 1-A^{2}\left[ n_{r}+A\right] ^{-2}\right\}
\medskip $ \\ 
$\lambda _{-,q=-1}=B^{2}\left\{ 1-A^{2}\left[ n_{r}-A+1\right] ^{-2}\right\}
\medskip $%
\end{tabular}%
\right.
\end{equation}%
Yet, it should be noted here that both even and odd quasi-parity eigenvalues 
$\lambda _{+,q=-1}$ and $\lambda _{-,q=+1}$ allow zero-modes (i.e.,
zero-energy) at $n_{r}=0$ and therefore they do not share the same spectrum
(a Nogami's and Toyama's [7] observation). However, $\lambda _{+,q=+1}$ and $%
\lambda _{-,q=+1}$ do not allow zero-modes and hence they have identical
spectra but with a missing lowest state in $\lambda _{-,q=+1}$. Similar
trend is also observed for $\lambda _{+,q=-1}$ and $\lambda _{-,q=-1}$ where
the lowest state is missed in $\lambda _{+,q=-1}$. Consequently, the
corresponding Dirac spectra read%
\begin{eqnarray}
E_{+,q} &=&\left\{ 
\begin{tabular}{l}
$E_{+,q=+1}=+\sqrt{m^{2}+B^{2}\left\{ 1-A^{2}\left[ n_{r}+A+1\right]
^{-2}\right\} }\medskip $ \\ 
$E_{+,q=-1}=+\sqrt{m^{2}+B^{2}\left\{ 1-A^{2}\left[ n_{r}-A\right]
^{-2}\right\} }\medskip $%
\end{tabular}%
\right. \\
E_{-,q} &=&\left\{ 
\begin{tabular}{l}
$E_{-,q=+1}=-\sqrt{m^{2}+B^{2}\left\{ 1-A^{2}\left[ n_{r}+A\right]
^{-2}\right\} }\medskip $ \\ 
$E_{-,q=-1}=-\sqrt{m^{2}+B^{2}\left\{ 1-A^{2}\left[ n_{r}-A+1\right]
^{-2}\right\} }\medskip $%
\end{tabular}%
\right.
\end{eqnarray}%
Evidently, energy-levels crossings obtain between the two sets of energies
in (51) or between the two sets of energies in (52). That is, a state $%
E_{+,q=+1}\left( n_{r}=n_{r1}\right) $ crosses with a state $%
E_{+,q=-1}\left( n_{r}=n_{r2}\right) $ when%
\begin{equation*}
n_{r1}+A+1=n_{r2}-A\Longrightarrow n_{r2}-n_{r1}=2A+1,
\end{equation*}%
and $E_{-,q=+1}\left( n_{r}=n_{r3}\right) $ crosses with $E_{-,q=-1}\left(
n_{r}=n_{r4}\right) $ when%
\begin{equation*}
n_{r3}+A=n_{r4}-A+1\Longrightarrow n_{r4}-n_{r3}=2A-1.
\end{equation*}%
Nevertheless, the energy sets in (51) and (52) loose their reality and
become pure imaginary in the following manner: 
\begin{eqnarray*}
E_{+,q=+1} &\in &%
\mathbb{C}
\text{ for }m^{2}+B^{2}<A^{2}B^{2}/\left( n_{r}+A+1\right) ^{2}, \\
E_{+,q=-1} &\in &%
\mathbb{C}
\text{ for }m^{2}+B^{2}<A^{2}B^{2}/\left( n_{r}-A\right) ^{2}, \\
E_{-,q=+1} &\in &%
\mathbb{C}
\text{ for }m^{2}+B^{2}<A^{2}B^{2}/\left( n_{r}+A\right) ^{2}, \\
E_{-,q=-1} &\in &%
\mathbb{C}
\text{ for }m^{2}+B^{2}<A^{2}B^{2}/\left( n_{r}-A+1\right) ^{2}.
\end{eqnarray*}

\section{Conclusion}

The inspiration of the current work is stimulated by our subsequent study
[20] of the hidden/built-in supersymmetric quantum mechanical language
and/or quasi-parity signatures on the spectrum of the (1+1)-Dirac equation.
However, as long as Dirac and Klein-Gordon wave equations are concerned, the
energy-levels crossing phenomenon/paradox (the discussion of which already
lies far beyond our current proposal, cf., e.g., [19] for more details) as a
spectral property or as a consequence of the supersymmetric language and/or
quasi-parity is left an \emph{almost-forgotten} one. Our purpose, even with
the current over-simplified radial Dirac/Klein-Gordon examples, was to fill
this gap at least partially.

In the light of the current study, we have observed that under different
settings of the magnetic interaction field $A\left( r\right) $, and likewise
the related interactions' way-of-coupling into Dirac equation, the two
ultimate/effective descendents, Dirac-Coulomb and Dirac-oscillator, perform
energy-levels crossing at different conditions. Moreover, our observations
in section 5 on the spectral properties of the radial supersymmetric partner
Hamiltonians (i.e., $\lambda _{\pm ,q=\pm 1}$ in (39), (40), (49), and (50))
re-confirm Nogami's and Toyama's [7] ones on the (1+1)-dimensional Dirac
equation.

Finally, we contemplate that the variety of settings (presented in this
work) of the related interactions' way-of-coupling (i.e., Eqs. (15), (27),
and (35)) into the (3+1)-dimensional radial Dirac equation would enrich the
number of exactly/quasi-exactly/conditionally-exactly solvable Dirac models.
Yet, the solution of the most general radial case in (13) still resides in
the mathematically challenging Hermitian, non-Hermitian, and
pseudo-Hermitian [20,21,24-26] Dirac territories.

\end{document}